\documentclass[aps,draft]{revtex4}
\usepackage[dvips]{graphicx}
\begin{document}

\title{On Dust Charging Equation}
\author{ Nodar L.Tsintsadze and Levan N.Tsintsadze}
\affiliation{Department of Plasma Physics, E.Andronikashvili
Institute of Physics, Tbilisi, Georgia}

\date{\today}

\begin{abstract}
A general derivation of the charging equation of a dust grain is
presented, and indicated where and when it can be used. A problem
of linear fluctuations of charges on the surface of the dust grain
is discussed.
\end{abstract}


\maketitle

In recent years a huge number of works have been devoted to the
investigation of dusty plasma and a dust in plasmas, taking into
account the charging equations. However, to the best of our
knowledge, in the literature the derivation of the charging
equation is missing. Also there does not exist any work where the
definition of surface charge and total current of electrons and
ions is accurately given.

Another flaw arises in the literature in considerations of
fluctuations of the surface charge and the total currents. Namely,
an inappropriate equations have been used in the studies of
perturbations of the surface charge and the total currents
yielding the damping of perturbations in a collisionless plasmas,
which is physically incorrect. Thus, the general derivation of the
charging equation of dust grain and the accurate definition of the
surface charge and the total currents remain unsolved problems of
the high importance.

In this Letter, we derive the charging equation from the Maxwell's
equations, without any assumptions, and indicate the validity of
this equation, that is when and where it can be used. We discuss
about the surface charge and the total currents of electrons and
ions in order to set right the use of equations for them. We show
that a small perturbations of the number of charge on surface in
the collisionless plasma, if the Landau damping effect is ignored,
once appear they are undamped, because there is no mechanism of
it.

We remind here readers that in Electrodynamics charges are treated
as points and the charge of a particle is an invariant quantity,
that is it does not depend on the choice of reference system. This
means that, when one derives the continuity equation and the
equation of motion of charged particles by the Lagrange equations,
one should assume that the charge is constant. Otherwise the
Lorentz force will have quite another form, so that even for one
particle case the classical field theory will breakdown. To
elucidate this, we use the Lagrangian for the charge of dust grain
in an electromagnetic (EM) field \cite{lan}
\begin{eqnarray}
\label{lag}
L=-m_Dc^2\sqrt{1-v^2/c^2}+\frac{Z_D}{c}(\vec{A}\cdot\vec{v})-Z_D
\varphi \ ,
\end{eqnarray}
where $m_D$ is the mass of the dust grain, $\vec{A}$ is the three
dimensional vector potential of EM field and $\varphi$ is the
scalar potential.

With the Lagrangian (\ref{lag}) at hand, as is well known one can
simply obtain the equations of motion of a charge in EM field by
employing the Lagrange equation
\begin{eqnarray}
\label{lage} \frac{d}{dt}\frac{\partial L}{\partial\vec{v}}=\frac{
\partial L}{\partial\vec{r}} \ .
\end{eqnarray}
The result is
\begin{eqnarray}
\label{eqm}
\frac{d\vec{p}}{dt}=Z_D\left(\vec{E}+\frac{1}{c}\vec{v}\times
\vec{B}\right)+\vec{F} \ ,
\end{eqnarray}
where the charge $Z_D$ is supposed to vary in time and space, the
expression in bracket on the right-hand side is the usual Lorentz
force, $\vec{F}$ is the additional force due to the variation of
charge of the dust grain and has such form
\begin{eqnarray}
\label{adf} \vec{F}=\left(\vec{A}\cdot\vec{v}-\varphi\right)\nabla
Z_D-\frac{\vec{A}}{c}\left(\frac{\partial}{\partial
t}+\vec{v}\cdot\nabla\right)Z_D \ .
\end{eqnarray}
It is obvious that the additional force $\vec{F}$ is nonphysical
and it is not defined, because of violation of the Gauge
invariance. Therefore, one must assume that the total charge on
surface of the dust grain is constant in time and space.

Moreover Eqs. (\ref{lag}),(\ref{eqm}) and (\ref{adf}) indicate
that the Lagrangian used above does not exactly describe the
interaction of the changeable charge with EM field. Hence, for the
latter a new Lagrangian should be introduced, from which can be
derived equations of motion that meet the Gauge invariance.

The same is, when we want to get the Maxwell's equations defined
by a total action for particles plus field in the Gaussian system
of units, which has the form \cite{lan}
\begin{eqnarray}
\label{ac} S=-\sum_\alpha\int m_\alpha cdS_\alpha-\frac{1}{c^2}
\int A_\imath j^\imath d\Omega-\frac{1}{16\pi c}\int F_{ik}
F^{ik}d\Omega \ ,
\end{eqnarray}
where $\alpha$ denotes the particle species, $j^\imath$ is the
total current density and the other notation is standard.

We specifically note here that in Eq.(\ref{ac}) it is assumed that
the charge on any species is constant. Therefore, employing any
equation of the macroscopic electrodynamics one must suppose that
charges of dust grains are constant, i.e., we may conclude that
despite extensive theoretical efforts, there is still no
electrodynamics for the macroscopic bodies  with the changing
charge.

The question, which we discuss in the following, concerns only one
dust grain in a plasma. Let us start with the definition of
surface charge on the dust grain. As is well known
\cite{lanel},\cite{jac} the distribution of excess charge of a
conducting body lies entirely on the surface of a conductor. Since
$E=0$ on the inner area of the surface, that is no electric filed
inside the grain, the tangential component of the electric field
at the surface is zero. But there is a normal component
$E_n=\vec{E}\cdot\vec{n}$ (where $\vec{n}$ is the unit vector) of
the electric field just outside the surface. The normal component
of the electric field takes very large values in the immediate
neighborhood of the surface. It is very important to emphasize
that the normal component $E_n$ pertains to the surface itself,
however it is not involved in the volume electrostatic problem,
because it falls off over distances comparable with the distances
between atoms.

Expression of the normal component $E_n$ can readily be obtained
by integration of the Poisson's equation. The result is
\begin{eqnarray}
\label{En} E_n=4\pi\int\rho dn=4\pi\sigma \ ,
\end{eqnarray}
where $\sigma$ and $\rho$ are the surface and the volume charge
densities, respectively.

Equation (\ref{En}) can be written in such form
\begin{eqnarray}
\label{sig} \sigma=-\frac{1}{4\pi}\frac{\partial\varphi}{\partial
n}\ ,
\end{eqnarray}
where the electrostatic potential $\varphi$ is constant on the
surface and fast decreases along the normal to the surface.

Using Eq.(\ref{sig}), we define the expression for the total
charge on the surface of one dust grain as
\begin{eqnarray}
\label{tch} Z_D=\int\sigma
dS=-\frac{1}{4\pi}\int\frac{\partial\varphi}{\partial n}dS\ ,
\end{eqnarray}
where the integral is taken over the whole surface, and $dS$ is
the element of area on the surface. Note that the equation
(\ref{tch}) reads just the Gauss theory. Namely, the Poisson's
equation in an integral form is
\begin{eqnarray}
\label{poi} \oint\vec{E}d\vec{S}=\oint(\vec{E}\cdot\vec{n})dS=
4\pi\int\sigma dS=4\pi Z_D \ .
\end{eqnarray}

We now derive the charging equation from the Maxwell's equation
\begin{eqnarray}
\label{max} curl\vec{B}=\frac{1}{c}\frac{\partial\vec{E}}{\partial
t}+\frac{4\pi}{c}\vec{j}\ ,
\end{eqnarray}
where $j$ is the total current density and can be expressed as
\begin{eqnarray}
\label{tcd} \vec{j}=\sum_\alpha e_\alpha n_\alpha \vec{v}_\alpha=
\sum_\alpha e_\alpha\int d^3v\vec{v}f_\alpha \ .
\end{eqnarray}
Integrating Eq.(\ref{max}) over surface of the grain and applying
Stoke's theorem
\begin{eqnarray*}
\int curl\vec{B}d\vec{S}=\oint\vec{B}d\vec{l}
\end{eqnarray*}
we get
\begin{eqnarray}
\label{magc}
\oint\vec{B}d\vec{l}=\frac{1}{c}\frac{\partial}{\partial t}\int
\vec{E}d\vec{S}+\frac{4\pi}{c}\int\vec{j}d\vec{S} \ .
\end{eqnarray}

Noting that the circulation of magnetic filed around any contour
for the closed surface equals zero, and using the definition
(\ref{tch}),(\ref{poi}) of the total charge on the surface of one
grain, we finally obtain the charging equation in the form
\begin{eqnarray}
\label{che} \frac{\partial Z_D}{\partial t}=-\sum_\alpha\int
\vec{j}_\alpha
d\vec{S}=-\sum_\alpha\int(\vec{j}_\alpha\cdot\vec{n})dS=
-\sum_\alpha I_\alpha \ ,
\end{eqnarray}
where $I_\alpha$ is the total current of electrons and ions.
Vector $d\vec{S}$ is  directed, as always, along the outward
normal to the surface of the grain, that is along the normal
towards the outside of the volume under consideration. It should
be emphasized that a left-hand side of Eq.(\ref{che}) is the
partial derivative in time, but not the total derivative with
respect to time.

Note that the equation (\ref{che}) is the exact expression of the
total charge on the surface of one grain for the case, when the
surface tension \cite{lanflu},\cite{lanstat} is not taken into
account. Detailed analysis of a charged surface is given in
Ref.\cite{ltsin}.

We now analyze the charging equation and explain how one should
understand the total current. In Eq. (\ref{che}) the current
density has the normal component $j_n$ alone, and differs from
zero even at Maxwell-Boltzmann distribution function. However,
well away from the grain $j_n=0$, the equation (\ref{che})
illustrates just one thing that the total number of charges on the
surface can change only if the total current of electrons $I_e$ is
less or more than the total current of ions $I_i$. Such situation
we have during charging processes of surface and this means that
the equation (\ref{che}) can be used for the charged surface
alone.

Let us in more detail consider the total current of particles,
which is
\begin{eqnarray}
\label{cur} I_\alpha=\sum_\alpha\int d\vec{S}\vec{j}_\alpha =\int
dS(\vec{n}\cdot\vec{j})=e_\alpha\int dS\int d^3vv_nf_\alpha \ .
\end{eqnarray}
The distribution function of particles $f_\alpha$ in general is
function of all space coordinates, time and momentum, i.e.,
$f_\alpha(\vec{p},\vec{r},t)\rightarrow
f(\vec{p},\varphi(\vec{r},t),\vec{A}(\vec{r},t))$. It follows that
in order to define the total current, one must integrate
right-hand side of Eq.(\ref{cur}) over the whole surface. Above
explanation about the expression (\ref{cur}) allows us to conclude
that the orbit-limited motion (OLM) approximation \cite{al} is
valid on the surface of dust grain, where the total current of
electrons and ions balance each other.

We next discuss the question of linearization of the charging
equation, small perturbations of the surface charges and damping
of these perturbations. We specifically note here that too many
publications were devoted to the study of damping of various
plasma waves in a collisionless plasmas (ignoring the Landau
damping) due to the fluctuation of the total surface charge. But
nobody discussed the mechanism driving to the damping. As we
already mentioned the charging equation describes any processes on
the surface of one grain. Equation (\ref{che}) exhibits that the
total surface charge changes in time (increases or decreases on
surface) depending on $I_e<I_i$ or $I_e>I_i$. So that we have
strictly nonstationary processes. Note that if we evoke a small
perturbation $Z_D=Z_{D0}+\delta Z_D$, then the average $<\delta
Z_D>=0$ and the total charge $<Z_D>$ remains constant. If $\delta
Z_D$ is considered in linear approximation, then once appear, it
will remain on the surface without damping. In this case one can
talk about the surface waves on the surface of one grain
\cite{ltsin}. As is well known the surface waves in the
collisionless medium are undamped.

As an example, we will show by a simple model (without losing
generality) that the fluctuation of the total surface charge
cannot lead to damping of the surface waves. To this end, we write
the variation of the total current $I=I_e+I_i$ in such form
\begin{eqnarray}
\label{vtc} \delta I=\sum_\alpha\delta
I_\alpha=\sum_\alpha\delta\int
dS(\vec{n}\cdot\vec{j}_\alpha)\approx\sum_\alpha 4\pi r_g^2\delta
j_{n\alpha}=4\pi r_g^2\sum_\alpha e_\alpha n_{0\alpha}v_{n\alpha}
\ ,
\end{eqnarray}
where $r_g$ is the radius of dust grain, $n_{0\alpha}$ is the
equilibrium density. Since $v_{n\alpha}$ is the normal component
of velocity, then $v_{n\alpha}=\frac{\partial\delta
r_{n\alpha}}{\partial t}$. So, the equation (\ref{che}) for the
perturbation reads $\frac{\partial\delta Z_D}{\partial
t}=-\sum_\alpha\beta_\alpha\frac{\partial\delta
r_{n\alpha}}{\partial t}$ or
\begin{eqnarray}
\label{ptc} \delta Z_D=-\sum_\alpha\beta_\alpha\delta r_{n\alpha}
\ ,
\end{eqnarray}
where $\beta_\alpha=4\pi r_g^2e_\alpha n_{0\alpha}$ is a constant,
and $\delta r_{n\alpha}$ is proportional to the variation of the
surface charge $\sigma$.

The equation of motion for the normal component of velocity on the
surface can be written as
\begin{eqnarray}
\label{eqnv} \frac{\partial^2\delta r_{n\alpha}}{\partial
t^2}=\frac{e_\alpha E_n}{m_\alpha} \ .
\end{eqnarray}

Use of Eq.(\ref{En}) in Eq.(\ref{eqnv}) for the linear
perturbations $\delta\sigma\sim e^{-i\omega t}$ yields the
solution
\begin{eqnarray}
\label{sol} \delta r_{n\alpha}=-\frac{4\pi
e_\alpha}{m_\alpha\omega^2}\delta\sigma \ .
\end{eqnarray}
Substituting Eq.(\ref{sol}) into Eq.(\ref{ptc}) we finally obtain
the following expression for the variation of the total charge on
the surface
\begin{eqnarray*}
\delta Z_D=\frac{\sum_\alpha\omega_{p\alpha}^2}{\omega^2}4\pi
r_g^2\delta\sigma \ ,
\end{eqnarray*}
where $\omega_{p\alpha}$ is the plasma frequency of the particle
species $\alpha$. Note that the same result we can get from the
kinetic equation if we neglect the Landau damping effect.

In summary, we have derived the charging equation of a dust grain
immersed in a plasma and have shown when and where one can use it.
We have discussed the electrodynamic properties of the dusty
plasma, and concluded that a charge on the dust grain must be
considered as a point particle with the constant charge, because
today does not exist the electrodynamics of medium with variable
charge of particles.

\end{document}